\documentclass{jfm}
\usepackage{amsmath}
\usepackage{amssymb}
\usepackage{lmodern}
\usepackage{color}
\definecolor{red}{rgb}{1,0,0}

\usepackage[T1]{fontenc}
\usepackage[utf8]{inputenc}
\usepackage{graphicx}
\usepackage{dcolumn}
\usepackage{bm}
\usepackage{hyperref}

\newcommand{\beq}{\begin{equation}}
\newcommand{\eeq}{\end{equation}}
\newcommand{\bal}{\begin{align}}
\newcommand{\eal}{\end{align}}

\shorttitle{NLS equation for gravity-capillary waves with shear}
\shortauthor{H.-C. Hsu, C. Kharif, M. Abid and Y.-Y. Chen}
\title{A nonlinear Schr\"{o}dinger equation for gravity-capillary water waves on arbitrary depth with constant vorticity: Part I}
\author{H.-C. Hsu\aff{1}, C. Kharif\aff{2}\corresp{\email{kharif@irphe.....}}, M. Abid\aff{2} \and Y.-Y. Chen\aff{1}}
\affiliation{\aff{1}Department of Marine Environment and Engineering, National Sun Yat-sen University, Kaohsiung, 801 Taiwan.
\aff{2}Aix Marseille Universit\'e, CNRS, Centrale Marseille, IRPHE UMR 7342, 13384, Marseille, France}

\begin{document}
\maketitle
\begin{abstract}
A nonlinear Schr\"{o}dinger equation for the envelope of two-dimensional gravity-capillary waves propagating at the free surface of a vertically sheared current of constant vorticity is derived. In this paper we extend to gravity-capillary wave trains the results of \citet{thomas2012pof} and complete the stability analysis and stability diagram of \citet{Djordjevic1977} in the presence of vorticity. Vorticity effect on the modulational instability of weakly nonlinear gravity-capillary wave packets is investigated. It is shown that the vorticity modifies significantly the modulational instability of gravity-capillary wave trains, namely the growth rate and instability bandwidth. It is found that the rate of growth of modulational instability of short gravity waves influenced by surface tension behaves like pure gravity waves: (i) in infinite depth, the growth rate is reduced in the presence of positive vorticity and amplified in the presence of negative vorticity, (ii) in finite depth, it is reduced when the vorticity is positive and amplified and finally reduced when the vorticity is negative. The combined effect of vorticity and surface tension is to increase the rate of growth of modulational instability of short gravity waves influenced by surface tension, namely when the vorticity is negative. The rate of growth of modulational instability of capillary waves is amplified by negative vorticity and attenuated by positive vorticity. Stability diagrams are plotted and it is shown that they are significantly modified by the introduction of the vorticity. 
\vspace{0.1cm}
\end{abstract}

\section{Introduction}
Generally, gravity-capillary waves are produced by wind which generates firstly a shear flow in the uppermost layer of the water and consequently these waves propagate in the presence of vorticity. These short waves play an important role in the initial development of wind waves, contribute to some extent to the sea surface stress and consequently participate in air-sea momentum transfer. Accurate representation of the surface stress is important in modelling and forecasting ocean wave dynamics. Furthermore, the knowledge of their dynamics at the sea surface is crucial for satellite remote sensing applications. 
\newline
In this paper we consider both the effect of surface tension and vorticity due to a vertically sheared current on the modulational instability of a weakly nonlinear periodic short wave trains. Recently, \citet{thomas2012pof} have derived a nonlinear Schr\"{o}dinger equation for pure gravity water waves on finite depth with constant vorticity. Their main findings were (i) a restabilisation of the modulational instability for waves propagating in the presence of positve vorticity whatever the depth and (ii) the importance of the nonlinear coupling between the mean flow induced by the modulation and the vorticity. One of our aim is to extend Thomas' investigation to the case of gravity-capillary waves propagating on a vertically sheared current.
\vspace{0.1cm}
\newline
The number of studies on the computation of steadily propagating periodic gravity waves on a vertically sheared current is important. For a review one can refer to the paper by \citet{thomas2012pof}. On the opposite, investigations devoted to the calculation of gravity-capillary waves in the presence of horizontal vorticity is rather meagre. One can cite \citet{Bratenberg1993} who used a third-order Stokes expansion for periodic gravity-capillary waves travelling on an opposing current and \citet{Hsu2016} who extended this work to the case of co- and counter-propagating waves. \citet{Kang2000} computed periodic and solitary gravity-capillary waves in the presence of constant vorticity on finite depth. They derived analytical solutions for small amplitude waves and numerical solutions for steeper waves. \citet{Wahlen2006} proved the rigorous existence of periodic gravity-capillary waves in the presence of constant vorticity.
\vspace{0.1cm}
\newline
To our knowledge, the unique study concerning the modulational instability of gravity-capillary waves travelling on a verticaly sheared current is that of \citet{Hur2017}. The stability of irrotational gravity-capillary waves has been deeply investigated by several authors. \citet{Djordjevic1977} and \citet{Hogan1985} derived nonlinear envelope equations and considered the modulational instability of periodic gravity-capillary waves. Note that in the gravity-capillary range, three-wave interaction is possible whereas modulational instability corresponds to a four-wave resonant interaction. The numerical computations were extended to capillary waves by \citet{Chen1985} and \citet{Tiron2012}. 
\citet{Zhang1986} investigated numerically the stability of gravity-capillary waves including, besides the four-wave resonant interaction, three-wave and five-wave resonant interactions. For a review on stability of irrotational gravity-capillary, one can refer to the review paper by \citet{Dias1999}.
\vspace{0.1cm}
\newline
This study is devoted to the modulational instability of weakly nonlinear gravity-capillary wave packets propagating at the surface of a vertically sheared current of finite depth. In section 2, the governing equation are given and the nonlinear Schr\"{o}dinger equation in the presence of surface tension and constant vorticity is derived by using a multiple scale method. In section 3, the linear stability analysis of a weakly nonlinear wave train is carried out as a function of the Bond number, the dispersive parameter and the intensity of the vertically sheared current. 

\section{Derivation of the NLS equation in the presence of surface tension and vorticity}
\begin{figure}
\begin{center}
 \includegraphics[width=0.7\linewidth]{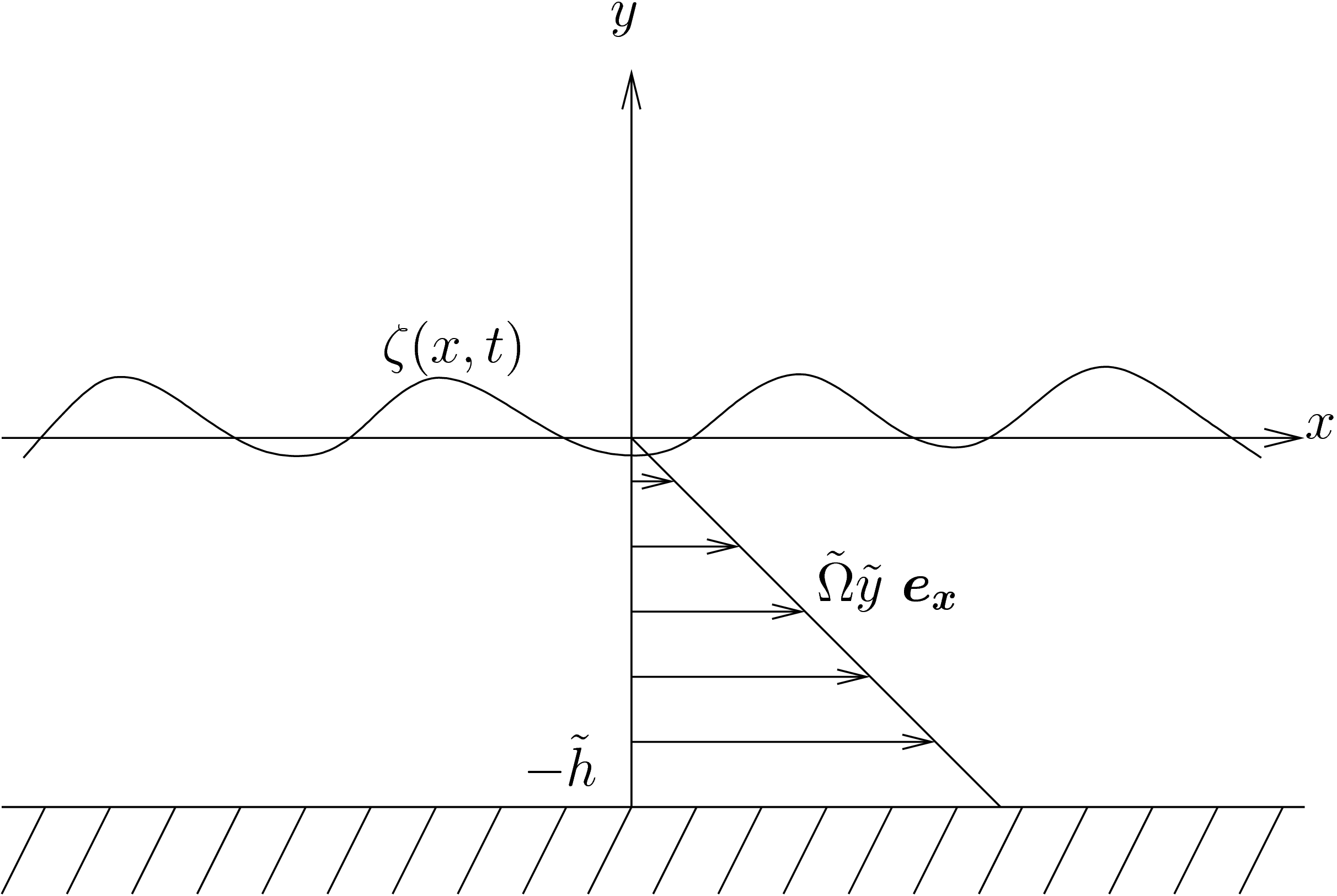}
 \end{center}
 \caption{Sketch of the two-dimensional flow.}
 \label{fig:basic}
\end{figure}

We consider the  modulational instability of weakly nonlinear surface gravity-capillary wave trains in the presence of vorticity. Our investigation is confined to  two-dimensional water waves propagating in finite depth. 
Viscosity is disregarded and the fluid is considered incompressible. The geometry configuration is presented in figure \ref{fig:basic}.

We choose an eulerian frame $(Oxyz)$ with unit vectors $(\vec{e}_x,\vec{e}_y,\vec{e}_z)$. The vector $\vec{e}_y$ is oriented upwards so that the gravity is $\vec{g}=-g\vec{e}_y$ with $g>0$. The equation of the undisturbed free surface is $y=0$ whereas the disturbed free surface is  $y=\zeta(x,t)$. The bottom is located at $y=-h$.
\newline
The waves are travelling at the surface of a vertically sheared current of constant vorticity. We consider an underlying current given by $\vec{u}_0=\Omega y \vec{e}_x$,
so that the fluid velocity reads
\beq
\vec{u}=\vec{u}_0+\vec{\nabla}\phi,
\eeq
where $\nabla\phi(x,y,z,t)$ is the wave induced velocity. The waves are potential due to the Kelvin theorem which states that vorticity is conserved for a two-dimensional flow of an incompressible and inviscid fluid with external forces deriving from a potential.
\newline
The potential $\phi$ satisfies the Laplace equation
\begin{equation}
\nabla^2\phi=0,
\end{equation}
and the Euler equation can be written as follows
\begin{equation}
\vec{\nabla}(\phi_t+\frac{1}{2}u^2+\frac{P}{\rho_w}+gy)=\vec{u}\wedge\vec{\omega},
\label{Euler0}
\end{equation}
 with $\vec{\omega}$ the vorticity vector along $z$, $P$ the pressure and $\rho_w$ the water density. Subscripts stand for derivatives in corresponding variables. 
\vspace{0.1cm}
\newline
Using the Cauchy-Riemann relations
\begin{equation}
\psi_y=\phi_x,\quad\psi_x=-\phi_y
\label{Cauchy-Riemann}
\end{equation}
where $\psi$ is the stream function
\begin{equation}
\vec{u}\wedge\vec{\omega}=\vec{\nabla}(\frac{1}{2}\Omega^2 y^2 + \Omega \psi)
\end{equation}
The Euler  equation (\ref{Euler0}) can be rewritten as follows
\begin{equation}
\vec{\nabla}(\phi_t+\frac{1}{2}\phi_x^2+\frac{1}{2}\phi_y^2+\Omega y\phi_x+gy-\Omega\psi+\frac{P}{\rho_w})=0
\end{equation}
Spatial integration gives the Bernoulli equation
\begin{equation}
\phi_t+\frac{1}{2}\phi_x^2+\frac{1}{2}\phi_y^2+\Omega y\phi_x+gy-\Omega\psi+\frac{P}{\rho_w}=C(t)
\label{bernoulli}
\end{equation}
In the presence of surface tension, $T$, at the free surface $y=\zeta(x,t)$ the Laplace law writes
\begin{equation}
P=P_a-T \frac{\zeta_{xx}}{(1+\zeta_x^2)^{3/2}}
\end{equation} 
where $P_a$ is the atmospheric pressure and $T$ surface tension.
\vspace{0.1cm} 
\newline
The dynamic boundary condition at the free surface $y=\zeta$ is
\begin{equation}
\phi_t+\frac{1}{2}\phi_x^2+\frac{1}{2}\phi_y^2+\Omega \zeta\phi_x+g\zeta-\Omega\psi-\frac{T}{\rho_w} \frac{\zeta_{xx}}{(1+\zeta_x^2)^{3/2}}=0
\label{Bernoulli}
\end{equation}
Witout loss of generality, we set $P_a=0$ and incorporate $C(t)$ into the potential $\phi$.
\vspace{0.1cm}
\newline
Along with these, we have the kinematic free surface boundary condition
\beq
\zeta_t+\zeta_x(\phi_x+\Omega y)-\phi_y=0,\quad y=\zeta(x,t)
\eeq
and the bottom boundary condition
\beq
\phi_y=0,\quad y=-h
\eeq
Following \citet{thomas2012pof} we can remove $\psi$ by deriving (\ref{Bernoulli}) with respect to $x$ and then using relations (\ref{Cauchy-Riemann}) , keeping in mind that we are dealing with low-steepness waves, and that (\ref{Bernoulli}) is evaluated in $y=\zeta$, we get the equation

\begin{align}
\phi_{tx}&+\phi_{ty}\zeta_x+\phi_x(\phi_{xx}+\phi_{xy}\zeta_x)+\phi_y(\phi_{xy}+\phi_{yy}\zeta_x)+\Omega\zeta_x\phi_x\nonumber\\
&+\Omega \zeta(\phi_{xx}+\phi_{xy}\zeta_x)+g\zeta_x+\Omega(\phi_y-\phi_x\zeta_x)\nonumber\\
&-\frac{T}{\rho_w}(\zeta_{xxx}-\frac{3}{2}\zeta_x^2\zeta_{xxx}-3\zeta_{xx}^2\zeta_x)=0,\enskip y=\zeta(x,t),\label{DFSBC}
\end{align}
that matches the one first derived in \citet{thomas2012pof} for $T=0$.
\vspace{0.1cm}
\newline
Following \citep{Davey1974}, we look for solutions depending on slow variables $(\xi,\tau) = (\varepsilon (x-c_g t),\varepsilon^2 t)$ where $\varepsilon=ak$ ($\varepsilon \ll 1$) and $a$,$k$ and $c_g$ are the amplitude, wavenumber and group velocity of the carrier wave, respectively. The system of governing equations becomes
\beq
\varepsilon^2 \phi_{\xi\xi}+\phi_{yy} = 0,\quad-h \leq y \leq \zeta(\xi,\tau),\label{laplace2}
\eeq

\beq
\phi_y=0,\quad y=-h,\label{bottom2}
\eeq

\beq
\varepsilon^2 \zeta_{\tau}-\varepsilon c_g \zeta_{\xi}+\varepsilon \zeta_{\xi}(\varepsilon \phi_{\xi}+\Omega[y+h])-\phi_y=0,\quad y=\zeta(\xi,\tau),\label{KFSBC2}
\eeq
\begin{align}
\varepsilon^3 \phi_{\tau \xi}&-\varepsilon^2 c_g \zeta_{\xi}+\varepsilon^3 \phi_{\tau y}\zeta_{\xi}-\varepsilon^2 c_g \phi_{\xi y} \zeta_{\xi}+\varepsilon^3 \phi_{\xi}(\phi_{\xi\xi}+\phi_{\xi y}\zeta_{\xi})\nonumber\\
&+\varepsilon \phi_y(\phi_{\xi y}+\phi_{yy}\zeta_{\xi})+\varepsilon^2\Omega\zeta_{\xi}\phi_{\xi}+\varepsilon^2 \Omega \zeta (\phi_{\xi \xi}+\phi_{\xi y}\zeta_{\xi})+\varepsilon g\zeta_{\xi}\nonumber\\
&+\Omega(\phi_y-\varepsilon^2 \phi_{\xi}\zeta_{\xi})-\varepsilon^3 \frac{T}{\rho_w}(\zeta_{\xi\xi\xi}-\frac{3}{2}\varepsilon^2 \zeta_{\xi}^2\zeta_{\xi\xi\xi}-3 \varepsilon^2 \zeta_{\xi\xi}^2\zeta_{\xi})=0,\quad y=\zeta(\xi,\tau)\label{DFSBC2}
\end{align}
An asymptotic solution to the system (\ref{laplace2}-\ref{bottom2}-\ref{KFSBC2}-\ref{DFSBC2}) is sought in the following form
\beq
\phi=\sum_{n=-\infty}^{+\infty}\phi_n E^n,\quad \zeta=\sum_{n=-\infty}^{+\infty}\zeta_n E^n,
\eeq
where $E=e^{i(kx-\omega t)}$ is a plane wave with $\omega$ the frequency of the carrier wave. We impose that $\phi_{-n}=\bar{\phi}_n$ and $\zeta_{-n}=\bar{\zeta}_n$ where the bar denotes complex conjugate, so that the functions are real. 
The amplitudes $\phi_n$ and $\zeta_n$ are then expanded in a perturbation series in terms of $\varepsilon=a k$
\beq
\phi_n=\sum_{j=n}^{+\infty}\varepsilon^j \phi_{nj},\quad\zeta_n=\sum_{j=n}^{+\infty}\varepsilon^j \zeta_{nj}. \label{perturbation}
\eeq
The terms depending on surface tension occur only at a higher order. 
The expansions (\ref{perturbation}) are substituted into the system of equations. The linear Laplace equation (\ref{laplace2}) is easier to handle, since solutions can be derived iteratively. 
Here we will simply write the first order solution for $\phi_{11}$, that is obtained by using the bottom boundary condition (\ref{bottom2})
\beq
\phi_{11}=A(\xi,\tau)\frac{\cosh[k(y+h)]}{\cosh(kh)},
\eeq
where the slow-varying function $A(\xi,\tau)$ will be used to express all other terms. Higher-order expansions of the Laplace equation introduce more unknown functions as solutions. Nevertheless, through expansions of the boundary conditions they can be all combined to $A(\xi,\tau)$.
\newline
The evolution of this unknown will depend on the initial condition $A(\xi,0)$. We then use (\ref{perturbation}) in the dynamic and kinematic free surface boundary conditions, and collect terms of equal power in $\varepsilon$ and $E$, which allows the expressions for the $\zeta_{ij}$ and $\phi_{ij}$ to be found successively.
\newline
The calculations are somewhat tedious but some steps are of interest. At first, the linear dispersion relation is derived

\beq
\omega^2 + \sigma \Omega \omega - \sigma g k(1+\kappa) =0,\label{lindisp}
\eeq
where $\sigma=\tanh(\mu)$ with $\mu=kh$ and $\kappa = \frac{{T} k^2}{\rho_w g}$.
\vspace{0.2cm}
\newline
The relation between $A(\xi,\tau)$ and $\zeta_{11}$ is the following
\beq
\zeta_{11}=i\frac{\omega (1+X)}{g(1+\kappa)} A(\xi,\tau),\label{potential_surface}
\eeq
where $X=\sigma \Omega/ \omega$
\newline
From the above dispersion relation we can show easily that $X>-1$. We note that $X$ depends also on the surface tension through $\omega$ and its associated dispersion relation.
It is also to be noted that the expression of the mean-flow term, which is important on the developement of the modulational instability, is similar to that of \citet{thomas2012pof}. Nevertheless, surface tension takes place through the phase velocity $c_p$, the group velocity $c_g$ and $\omega$.
\beq
(c_g (c_g+\Omega h)-gh)\phi_{01,\xi}=(\frac{g \sigma}{c_p^2} (2 \omega +\sigma \Omega)+ k^2 c_g (1-\sigma^2)) |A|^2,
\eeq
and
\beq
g\zeta_{02} = (c_g+\Omega h)\phi_{01,\xi}-k^2 (1-\sigma^2) |A|^2.
\eeq
Although the expressions are identical to those of \citet{thomas2012pof}, it should be noted that the surface tension acts through the dispersion relation, affecting $\omega$, $c_p$ and $c_g$. 
\newline
It is at the order $\mathcal{O}(\varepsilon^3 E)$ that the nonlinear Schrödinger equation is found for the potential envelope $A$, so that
\begin{align}
 i A_{\tau}+\alpha A_{\xi\xi}=\gamma |A|^2 A,\label{NLS}
\end{align}
where the coefficients depend on $(\kappa,\Omega,kh)$. 
\newline
Then the dispersion coefficient reads
\beq
\begin{aligned}
\alpha=\frac{-\omega}{k^2 \sigma(2+X)}&\left[ \sigma \rho^2 +\mu\frac{1+X}{1+\kappa}(\sigma[\sigma+\mu(1-\sigma^2)]-1)\right.\\
&\left.+\mu(1-\sigma^2)(\rho-\mu\sigma)X-\frac{\kappa}{1+\kappa}\alpha_1\right],\\
\end{aligned}
\eeq
with
\begin{align}
 \alpha_1=-\mu(1+X)(1-\sigma^2)(\mu\sigma-1)+\sigma(1+X)(1+2\rho)\\
 +2\left(\sigma\rho+\mu(1-\sigma^2)X-2\frac{\sigma\kappa}{1+\kappa}(1+X)\right),\nonumber
\end{align}
where $\rho=c_g/c_p$ is here the ratio of the group velocity to the phase velocity of the carrier. It can be expressed in a concise form
\beq
\rho=\frac{(1-\sigma^2)\mu+(1+X)(\sigma+\frac{2 \sigma \kappa}{1+\kappa})}{\sigma(2+X)},
\eeq
which depends only on $\mu,\kappa,X$.
The nonlinear coefficient is 

\begin{align}
\gamma=&\frac{k^4}{2\omega(1+X)(2+X)}\left[-\frac{3\sigma^2\kappa}{1+\kappa}(1+X)^2-2(1+\kappa)(1-\sigma^2)[(1+X)^2-\sigma^2]\right.\nonumber\\
&+\sigma^2 (1+X)(8+6X)+\frac{1+X}{\sigma^2-\kappa(3-\sigma^2+3 X)}\gamma_1\label{nl_coeff}\\
&+2\left.\frac{(1+X)(2+X)+\rho(1+\kappa)(1-\sigma^2)}{(1+\kappa)(\rho^2+\mu\rho\frac{X}{\sigma}-\frac{\mu(1+X)}{\sigma(1+\kappa)})}\gamma_2\right],\nonumber
\end{align}
with
\begin{align}
 \gamma_1&=9-10\sigma^2+\sigma^4+(18-4\sigma^2-4\sigma^4)X+(15+3\sigma^2)X^2\nonumber\\
 &+(6+2\sigma^2)X^3+X^4\\
 &+\kappa\left[ 21-10\sigma^2+\sigma^4+(42+2\sigma^2-4\sigma^4)X\right.\nonumber\\
&+\left.(30+12\sigma^2)X^2+(9+5\sigma^2)X^3+X^4\right],\nonumber
\end{align}
and finally
\begin{align}
 \gamma_2&=(1+\kappa)\left[(1+X)^2(1+\rho+\frac{\mu X}{\sigma})+1+X-\sigma(\rho\sigma+\mu X)\right]\nonumber\\
 &-\kappa(1+X)(2+X),
\end{align}
and we can check that these coefficients reduce to those of \citet{Djordjevic1977}, or \citet{Hogan1985} in deep water, if $\Omega=0$ and to those of \citet{thomas2012pof} if $\kappa=0$. 
\newline
The last term in brackets of equation (\ref{nl_coeff}) corresponds to the coupling between the mean flow due to the modulation and the vorticity which occurs at third-order. This coupling was found by \citet{thomas2012pof} for the case of pure gravity waves and has an important impact on the stability analysis of progressive wave trains.
\newline
We can see that in (\ref{nl_coeff}) there are two possible singularities that one should avoid, either
\beq
\sigma^2-\kappa(3-\sigma^2+3X)=0,
\eeq
which corresponds to the first gravity-capillary resonance $\kappa_c=\frac{\sigma^2}{3-\sigma^2}$ without vorticity,
or 
\beq
\rho^2+\rho\frac{\mu X}{\sigma}-\frac{\mu(1+X)}{\sigma(1+\kappa)}=0,
\eeq
which is rewritten as follows
\begin{equation}
c_g^2+\frac{g\mu}{\omega}X\frac{1+\kappa}{1+X}c_g-\frac{g^2\mu\sigma}{\omega^2}\frac{1+\kappa}{1+X}=0
\nonumber
\end{equation}
In the absence of vorticity, the latter condition reduces to $c_g^2=gh$ which matches the long wave - short wave resonance as shown by \citet{Davey1974} and \citet{Djordjevic1977}. In the presence of vorticity and for pure gravity waves the nonlinear coefficient becomes singular if the following condition is satisfied 
\begin{equation}
\{1+\frac{\mu \sigma\Omega(2+X)}{(1-\sigma^2) \mu +\sigma (1+X)}\} c_g^2 = gh
\nonumber
\end{equation}   
Note that this condition reduces to $c_g^2=gh$ in the absence of vorticity.

\section{Stability analysis and results}

\begin{figure}
\begin{center}
 \includegraphics[width=0.7\linewidth]{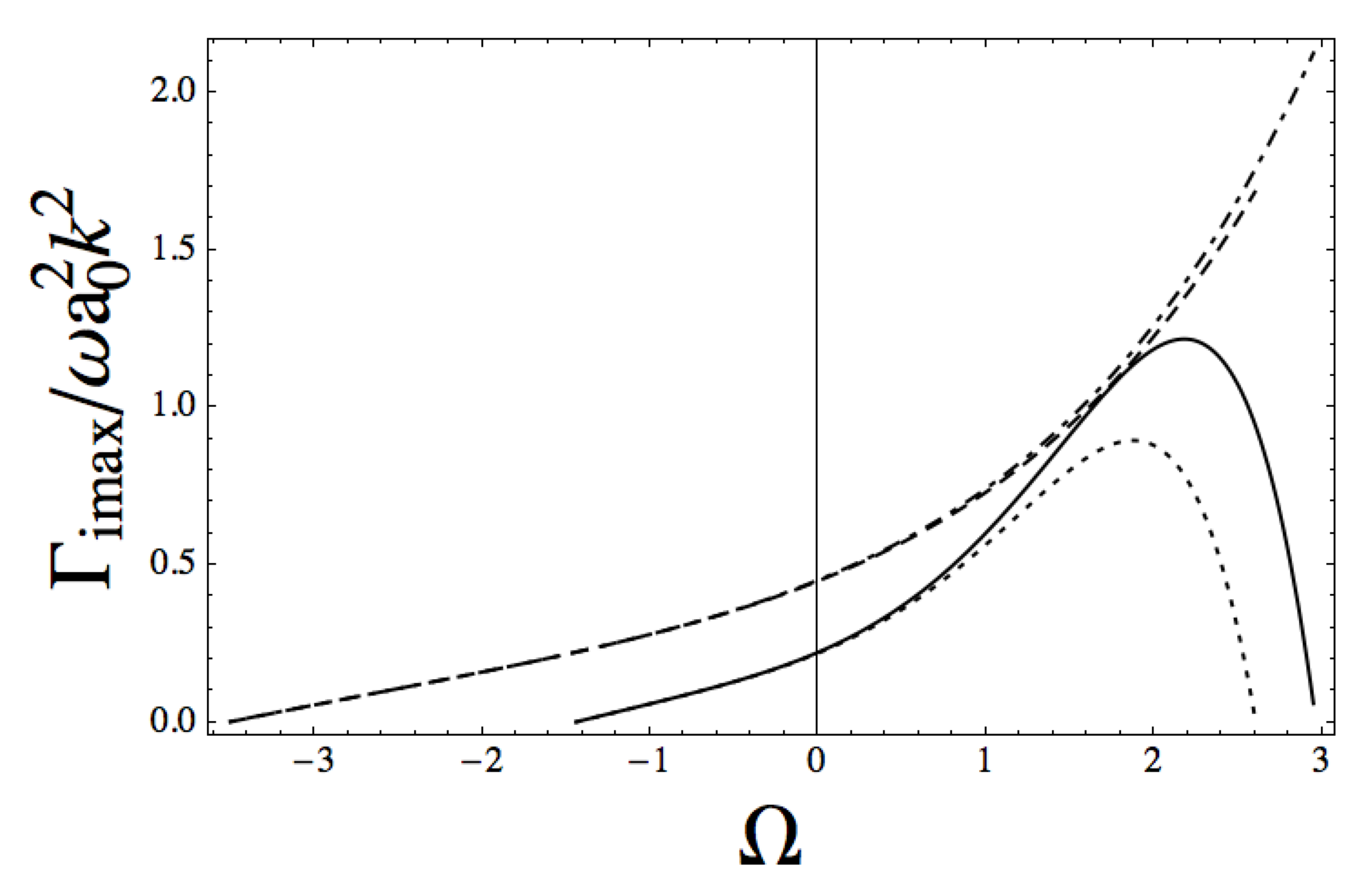}
 \end{center}
 \caption{Dimensionless maximal growth rate of modulational instability as a function of $\Omega$ in finite depth ($\mu=2$) and deep water ($\mu=\infty$). Solid line ($\kappa=0.005, \, \mu=2$); Dot-dashed line ($\kappa=0.005, \, \mu=\infty$) ; Dotted line ($\kappa=0, \, \mu=2$); Dashed line ($\kappa=0, \, \mu=\infty$)  }
 \label{Taux_max_0_005}
\end{figure}

\begin{figure}
\begin{center}
\includegraphics[width=0.7\linewidth]{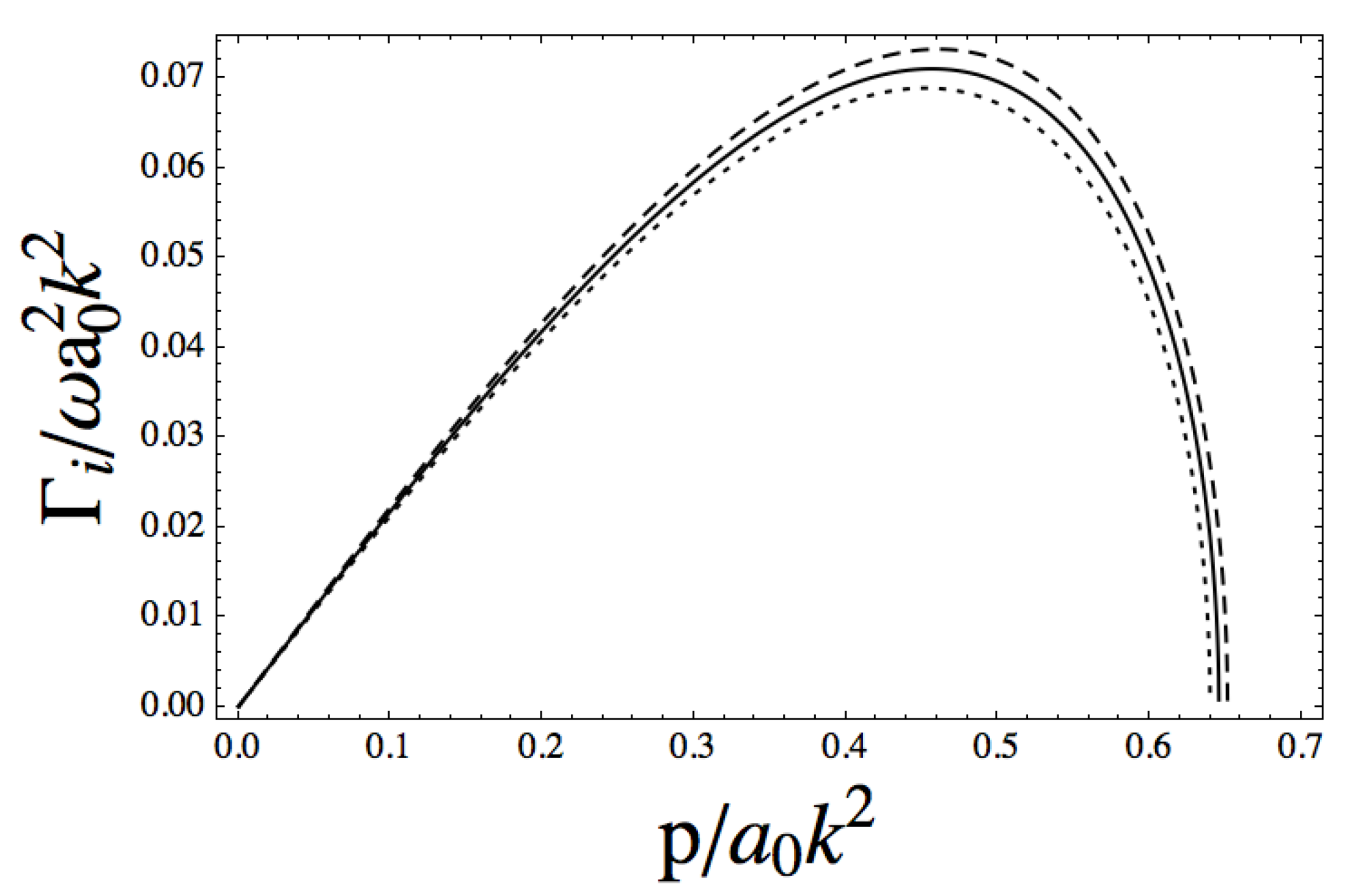}
\end{center}
\caption{Dimensionless growth rate of modulational instability of pure capillary waves in finite depth ($\mu=2$) as a function of the dimensionless wavenumber of the perturbation for several values of
$\Omega$. $\Omega=0$ (solid line);  $\Omega=2$ (dashed line); $\Omega=-2$ (dotted line).   }
\label{Taux_ka_infini_mu_2}
\end{figure}

Let us write $\zeta$ in the form
\[\zeta=\frac{1}{2}(\epsilon a e^{i(kx-\omega t)} + c.c.) + \mathcal{O}(\epsilon^2) \]
where $a=2 \zeta_{11}$ is the envelope of the free surface elevation and $c.c.$ denotes complex conjugation. Using (\ref{potential_surface}) the NLS equation (\ref{NLS}) is rewritten for the complex envelope $a(\xi,\tau)$ as follows

\beq
i a_{\tau} + \alpha a_{\xi\xi} = \tilde{\gamma} |a|^2 a, \label{amplitude_NLS}
\eeq
where
\[ \tilde{\gamma}=\frac{g^2}{4\omega^2} (\frac{1+\kappa}{1+X})^2 \gamma \]
\newline
The nonlinear coefficient $\tilde{\gamma}$ can be written in a more compact form
\[ \tilde{\gamma}=\frac{\omega^2}{4k^2\sigma^2} \gamma \]
\vspace{0.1cm}
\newline
In this section we consider the stability of a Stokes wave solution of the NLS equation (\ref{amplitude_NLS}) to infinitesimal disturbances. 
\newline
Equation (\ref{amplitude_NLS}) admits the following solution
\beq
a_s(\tau)=a_0 e^{-i \tilde{\gamma} a_0^2 \tau},
\eeq
with the initial condition $a_0$. 
\newline
We consider infinitesimal perturbations to this solution, in amplitude $\delta_a(\xi,\tau)$ and in phase $\delta_w(\xi,\tau))$, so that the perturbed 
solution $a_s'$ writes
\beq
a_s'=a_s(1+\delta_a)e^{i\delta_w},\label{Stokes_pert}
\eeq
Substituting this expression in the NLS equation (\ref{amplitude_NLS}), linearising and separating between real and imaginary parts, yields to a system of linear coupled partial differential equations with constant coefficients. Then, this system admits solutions of the form 
\begin{align}
 \delta_a&=\delta_{a_0} e^{i (p \xi -  \Gamma \tau)},\nonumber\\
 \delta_w&=\delta_{w_0} e^{i (p \xi -  \Gamma \tau)},
\end{align}
The necessary and sufficient condition for the existence of non-trivial solutions is 
\beq
\Gamma^2=\alpha p^2 (2\tilde{\gamma}a_0^2+\alpha p^2),
\eeq
The Stokes wave solution is stable when $\alpha(2\tilde{\gamma}a_0^2+\alpha p^2) \geq 0$ and unstable when $\alpha(2\tilde{\gamma}a_0^2+\alpha p^2) < 0$
\newline
The growth rate of instability is then
\begin{equation}
\Gamma_i=p (-2\tilde{\gamma} \alpha a_0^2-\alpha^2 p^2)^{1/2}
\nonumber
\end{equation}
We set $\alpha=\omega \alpha_2/k^2$ and $\tilde{\gamma}=\omega k^2 \tilde{\gamma_1}$, so that $\alpha_2$ and $\tilde{\gamma_1}$ are dimensionless functions of $\mu=kh$, $X=\sigma \Omega/\omega$ and $\kappa$ only. The growth rate of instability becomes
\begin{equation}
\Gamma_i=\frac{\omega \, p}{k^2} (-2\tilde{\gamma_1} \alpha_2  a_0^2 k^4 -\alpha_2^2 p^2)^{1/2}
\label{growthrate}
\end{equation}
The maximal growth rate is obtained for $p=\sqrt{-\tilde{\gamma_1}/\alpha_2}\, a_0 k^2$ and its expression is $\Gamma_{imax}=\sqrt{-\tilde{\gamma_1}/\alpha_2}\,
\sqrt{-\tilde{\gamma_1}\alpha_2}\,\omega(a_0 k)^2$. Note that instability occurs when $\tilde{\gamma_1}$ and $\alpha_2$ have opposite sign.
\newline
The growth rate of instability is written in the following dimensionless form
\begin{equation}
\frac{\Gamma_i}{\omega a_0^2 k^2}=\tilde{p} \,(-2 \tilde{\gamma_1}\alpha_2-\alpha_2^2 \tilde{p}^2 )^{1/2}
\label{growth_dimenensionless}
\end{equation}
where $\tilde{p}=p/(a_0 k^2)$
\newline
The dimensionless bandwidth of instability is $\Delta \tilde{p}=\sqrt{-2 \tilde{\gamma_1}/\alpha_2}$ and 
$\Delta p/k=\sqrt{-2 \tilde{\gamma_1}/\alpha_2} \, a_0 k$.
\vspace{0.2cm}
\newline
For $\kappa=0$ and $\Omega \neq 0$, equation (\ref{growth_dimenensionless}) gives the rate of growth of \citet{thomas2012pof}. In figure \ref{Taux_max_0_005} is plotted the dimensionless maximal growth rate of modulational instability of pure gravity waves and gravity waves influenced by surface tension effect ($\kappa=0.005$) as a function of $\Omega$ for infinite and finite depths. We can observe that combined effect of surface tension and vorticity increases significantly the rate of growth of the modulational instability of short gravity waves propagating in finite depth and in the presence of negative vorticity ($\Omega>0$) whereas the effect is insignificant in deep water. For positive vorticity ($\Omega<0$) the curves almost coincide in finite depth and deep water as well and the increase of the rate of growth due to surface tension is of order of $\kappa$.
\vspace{0.2cm}
\newline
For $\Omega=0$ and $\kappa \neq 0$, equation (2.20) of \citet{Djordjevic1977} becomes 
\begin{equation}
i a_{\tau}-\frac{\omega}{8k^2} \frac{1-6\kappa-3\kappa^2}{(1+\kappa)^2}a_{\xi\xi}=\frac{k^2\omega}{16}\frac{8+\kappa+2\kappa^2}{(1-2\kappa)(1+\kappa)}|a|^2a
\nonumber
\end{equation}
for the envelope of the surface elevation in deep water.
\newline
The coefficients $\tilde{\gamma_1}$ and $\alpha_2$ corresponding to this NLS equation are
\[\tilde{\gamma_1}=\frac{1}{16}\frac{8+\kappa+2\kappa^2}{(1-2\kappa)(1+\kappa)}, \qquad  \alpha_2=-\frac{1}{8} \frac{1-6\kappa-3\kappa^2}{(1+\kappa)^2}\]
Consequently, the rate of growth of modulational instability of pure capillary wave trains on infinite depth, obtained for $\kappa \rightarrow \infty$, is
\[\Gamma_i \rightarrow \frac{\omega}{8k^2}(3a_0^2 k^4 p^2-9p^4)^{1/2} \qquad \mathrm{as} \qquad \kappa \rightarrow \infty \]
which can be found in \citet{Chen1985}. The wavenumber of the fastest-growing modulational instability is $p_{\mathrm{max}}=a_0 k^2/\sqrt{6}$ and the maximum growth rate is $\omega (a_0 k)^2/16$. \citet{Tiron2012}  have extended the linear stability of finite-amplitude capillary waves on deep water subject to superharmonic and subharmonic perturbations without vorticity effect.
\newline
We have considered the case of pure capillary waves on deep water ($\kappa \rightarrow \infty$ and $\mu \rightarrow \infty$) in the presence of vorticity ($\Omega \neq 0$). The corresponding analytic expressions of $\tilde{\gamma_1}$ and $\alpha_2$ are
\begin{equation}
\tilde{\gamma_1}=-\frac{3+14X+23X^2+11X^3-3X^4}{24(X+1)(3X+2)}
\label{gamma1_cap}
\end{equation}
\begin{equation}
\alpha_2=\frac{3(X+1)(X^2+X+1)}{(2+X)^3}
\label{alpha2_cap}
\end{equation}
where $X=\Omega/\omega$ and $\omega=-\Omega/2 \pm \sqrt{(\Omega/2)^2 +k^3T/\rho_w}$.
\newline
Due to high wave frequency of capillaries on deep water we assume $\mid X \mid \ll 1$. The coefficients $\tilde{\gamma_1}$ and $\alpha_2$ becomes 
\begin{equation}
\tilde{\gamma_1}=-\frac{1}{16}(1+\frac{13}{6} X) + \mathcal{O}(X^2)
\label{gamma1_approx}
\end{equation}
\begin{equation}
\alpha_2=\frac{3}{8}(1+\frac{X}{2}) + \mathcal{O}(X^2)
\label{alpha2_approx}
\end{equation}
The rate of growth of modulational instability of capillary waves on deep water in the presence of vorticity is
\begin{equation}
\Gamma_i=\frac{\omega \, p}{8 k^2} \sqrt{3 a_0^2 k^4-9p^2 + (8a_0^2 k^4 - 9p^2) X} + \mathcal{O}(X^2)
\label{growth-cap}
\end{equation}
and in dimensionless form
\begin{equation}
\frac{\Gamma_i}{\omega \, a_0^2 k^2}=\frac{\tilde{p}}{8} \sqrt{3-9\tilde{p}^2 + (8-9 \tilde{p})X} + \mathcal{O}(X^2)
\label{growth_cap_adim}
\end{equation}
The maximal growth rate of instability is obtained for $p=(1+5X/6)a_0 k^2/\sqrt{6} + \mathcal{O}(X^2)$ and its value is $(1+13X/6)\omega a_0^2 k^2/16 +\mathcal{O}(X^2)$. The bandwidth of modulational instability is $\Delta p= (1+5X/6)a_0 k^2/\sqrt{3}$. 
\newline
Consequently, the rate of growth of modulational instability of capillary waves in deep water is larger for negative vorticity ($X>0$) than for positive vorticity ($X<0$). The bandwidth of instability presents the same trend.
\newline
In figure \ref{Taux_ka_infini_mu_2} is shown the dimensionless rate of growth of modulational instability of pure capillary waves in finite depth as a function of the wavenumber of the perturbation, for several values of $\Omega$. The rate of growth of instability increases as $\Omega$ increases as in infinite depth.
\begin{figure}
\begin{center}
 \includegraphics[width=0.6\linewidth]{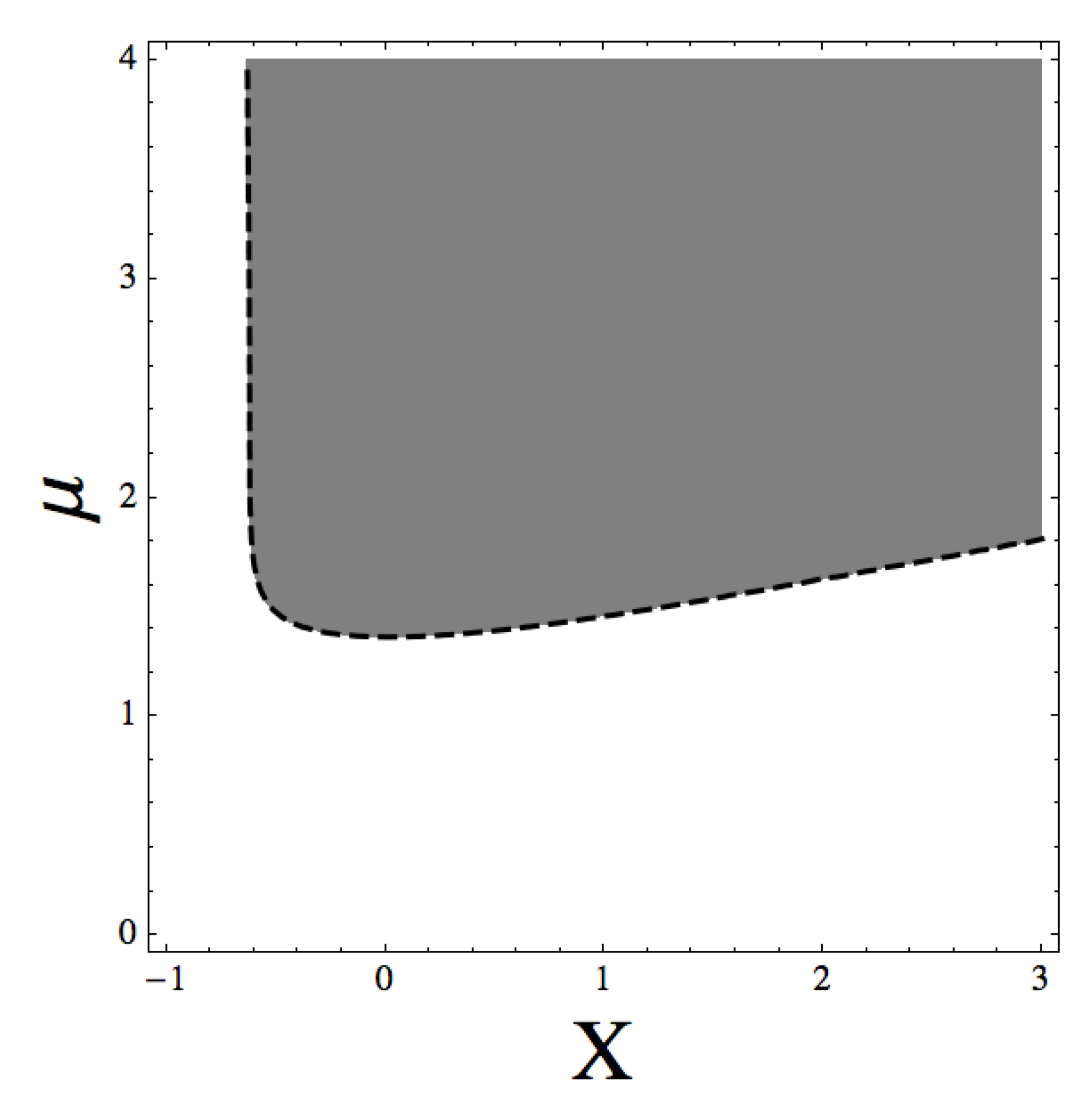}
 \end{center}
 \caption{$(\mu,X)$-instability diagram for gravity waves, matching the results of \citet{thomas2012pof} (dashed lines). Here, there is no surface tension.  The unstable regions are in gray whereas stable regions are in white. 
 For $X=0$ (or $\Omega=0$) the value $kh\approx1.363$ is found, below which there is no instability.}
 \label{fig:basic1}
\end{figure}
\begin{figure}
\begin{center}
 \includegraphics[width=0.6\linewidth]{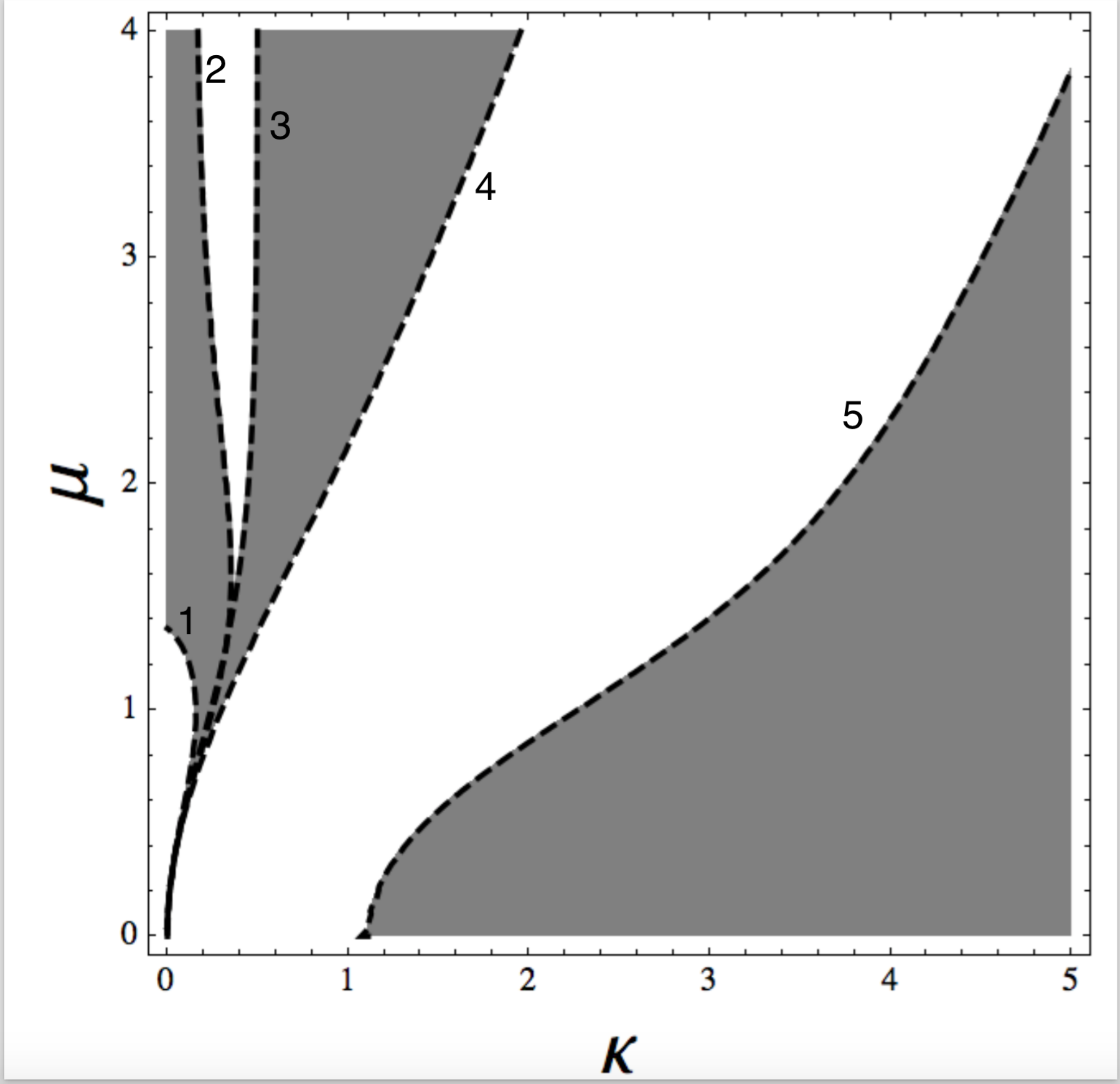}
 \end{center}
 \caption{$(\mu,\kappa)$-instability diagram for gravity-capillary waves, matching the results from \citet{Djordjevic1977} (dashed lines). Here, there is no vorticity. The unstable regions are in gray whereas stable regions are in white.}
 \label{fig:instab1}
\end{figure}
\vspace{0.2cm}
\newline
The sign of the product $\alpha\tilde{\gamma}$ determines the stability of the solution under infinitesimal perturbations. If the product is positive then the solutions are modulationally stable, otherwise they are modulationally unstable and grow exponentially with time. \citet{Davey1974} and \citet{Djordjevic1977} showed that this criterion which works for $1D$ propagation can be extended to the case of $2D$ propagation. In this way, our stability diagrams could be compared to those of \citet{Djordjevic1977} when $\Omega=0$. The linear stability analysis only captures the linear part of the instability, and thus its onset. We plot in the $(\mu=kh,\kappa)$-plane, for fixed values of the vorticity $\Omega$, the unstable and stable regions. As a check, the instability diagrams we obtain are compared in Figs. \ref{fig:basic1} and \ref{fig:instab1} with the same diagrams obtained by \citet{thomas2012pof} for $\kappa=0$ and \citet{Djordjevic1977} for $\Omega=0$. In that way, we can verify that these limiting cases are reproduced correctly. Following \citet{Djordjevic1977}, the boundaries of the unstable regions have been numbered from 1 to 5. Curve 1 crosses the $\mu$-axis at the point corresponding to restabilisation of the modulational instability. Note that this feature holds for two-dimensional water waves. Curve 2 corresponds to vanishing of the dispersive coefficient $\alpha$ and minimum phase velocity ($c_g=c_p$) whereas along curves 3 and 4 the nonlinear coefficient $\tilde{\gamma}$ is singular. These singularities define Wilton and long wave/short wave resonances, respectively. Curves 1 and 5 correspond to simple zeros of the nonlinear coefficient $\tilde{\gamma}$.
\newline
Curve 4 has the following asymptote
\begin{equation}
\mu=(1+\frac{\Omega^2}{2}-\sqrt{\frac{\Omega^2}{4}(4+\Omega^2)})(\frac{9}{4} \kappa-\frac{3}{4}), \qquad \mu \gg 1,
\nonumber 
\end{equation}
whereas curve 5 has the asymptote
\begin{equation}
\mu=\frac{9}{4}(1+\frac{\Omega^2}{2}-\sqrt{\frac{\Omega^2}{4}(4+\Omega^2)}\,)\, \kappa + \frac{1}{4}(-35+3\Omega^2 + \frac{29\Omega} { \sqrt{ 4+\Omega^2}}                                       
 -\frac{3\Omega^3} { \sqrt{ 4+\Omega^2} } )   \qquad \mu \gg 1,
\nonumber 
\end{equation}
For $\Omega=0$, the equations of \citet{Djordjevic1977} are redicovered except that instead of $-61/8$ we found $-35/4$ which is slightly different. The asymptotes have the same slope. In the region beteen these two asymptotes the capillary waves ($\kappa \gg 1$) are modulationally stable. This feature was emphasized by \citet{Djordjevic1977}  in the absence of vorticity.
\newline
In figures \ref{instab_omega_m1} to \ref{instab_omega_4} the effect of positive and negative vorticity on $(\mu=kh,\kappa)$ diagrams is investigated. The curves of \citet{Djordjevic1977} have been plotted to show the effect of the vorticity. As it can be observed, the vorticty has a significant effect on stability diagrams of gravity-capillary. Very recently, this feature was emphasized by \citet{Hur2017} who proposed a shallow water wave model with constant vorticity and surface tension, too. Although interesting this model suffers from shortcomings: (i) dispersion is introduced heuristically and is fully linear (ii) nonlinear terms due to surface tension effect are ignored (iii) the coupling between nonlinearity and dispersion is not taken into account.
\newline
As positive vorticity ($\Omega <0$) increases, we observe in figures \ref{instab_omega_m1}, \ref{instab_omega_m2}, \ref{instab_omega_m3} and \ref{instab_omega_m4} along the $\mu$-axis in the vicinity of $\kappa=0$ an increase of the region where the Stokes gravity-capillary wave train is modulationally stable. Consequently, gravity waves influenced by surface tension behave as pure gravity waves (see \citet{thomas2012pof}). Nevertheless, a very thin tongue of instability persists, near $\kappa=0$, in the shallow water regime. 
\newline\
As the intensity of negative vorticity ($\Omega>0$) increases the band of instability along the $\mu$-axis that corresponds to small values of $\kappa$ becomes narrower, as shown in figures \ref{instab_omega_1}, \ref{instab_omega_2}, \ref{instab_omega_3} and \ref{instab_omega_4}. Contrary to the case of positive vorticity, the region of restabilisation along the $\mu$-axis does not increase in the vicinity of $\kappa=0$.
\begin{figure}
\centering
\begin{minipage}{0.49\textwidth}
\centering
\includegraphics[width=\linewidth]{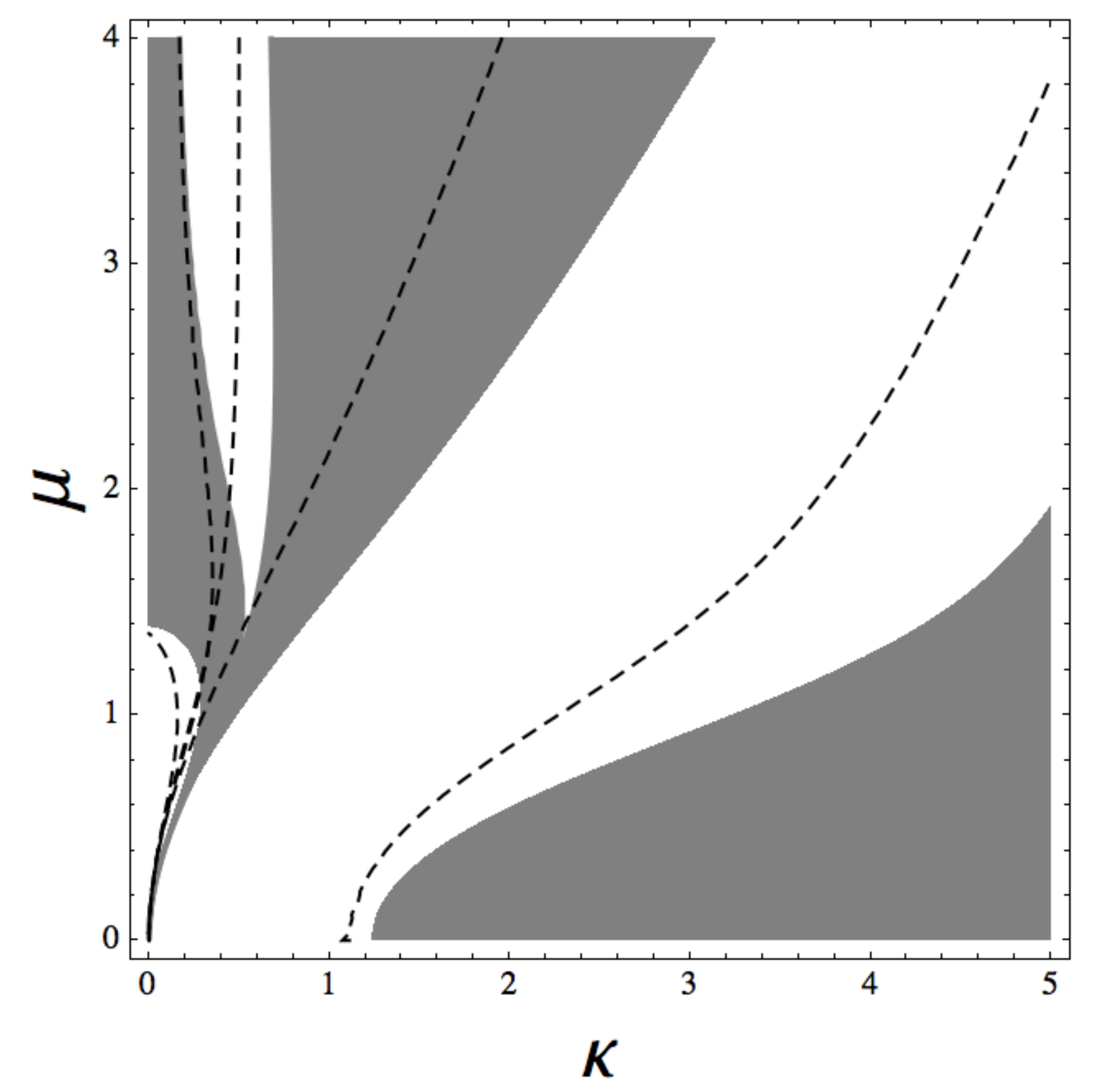} 
\caption{$(\mu,\kappa)$-instability diagram for $\Omega=-0.5$ (positive vorticity). The dashed lines correspond to $\Omega=0$.} 
\label{instab_omega_m1}
\end{minipage}\hfill
\begin{minipage}{0.49\textwidth}
\centering
 \includegraphics[width=\linewidth]{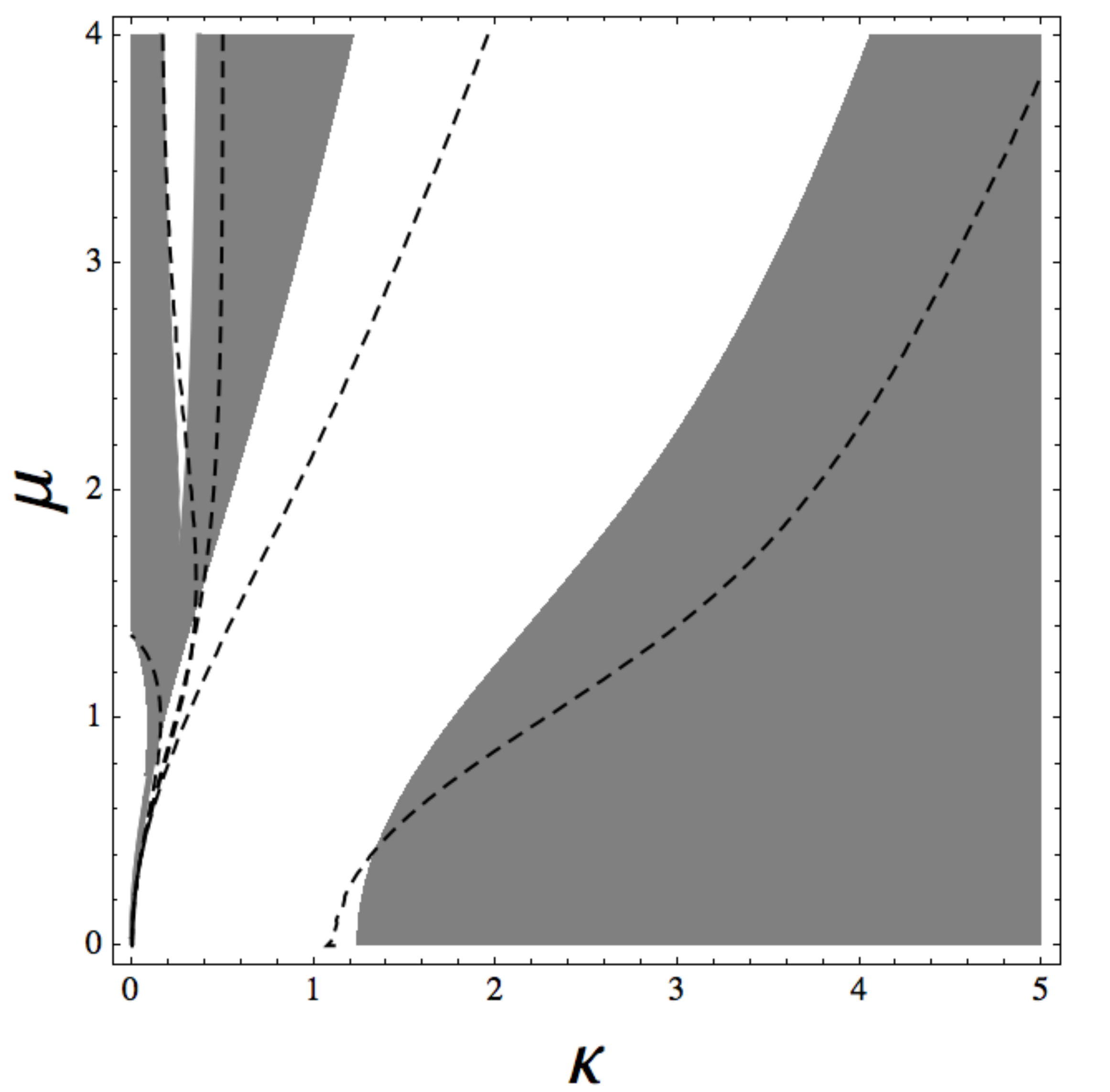}
\caption{$(\mu,\kappa)$-instability diagram for $\Omega=0.5$ (negative vorticity). The dashed lines correspond to $\Omega=0$.} 
\label{instab_omega_1}
\end{minipage}
\end{figure}
\begin{figure}
\centering
\begin{minipage}{0.49\textwidth}
\centering
\includegraphics[width=\linewidth]{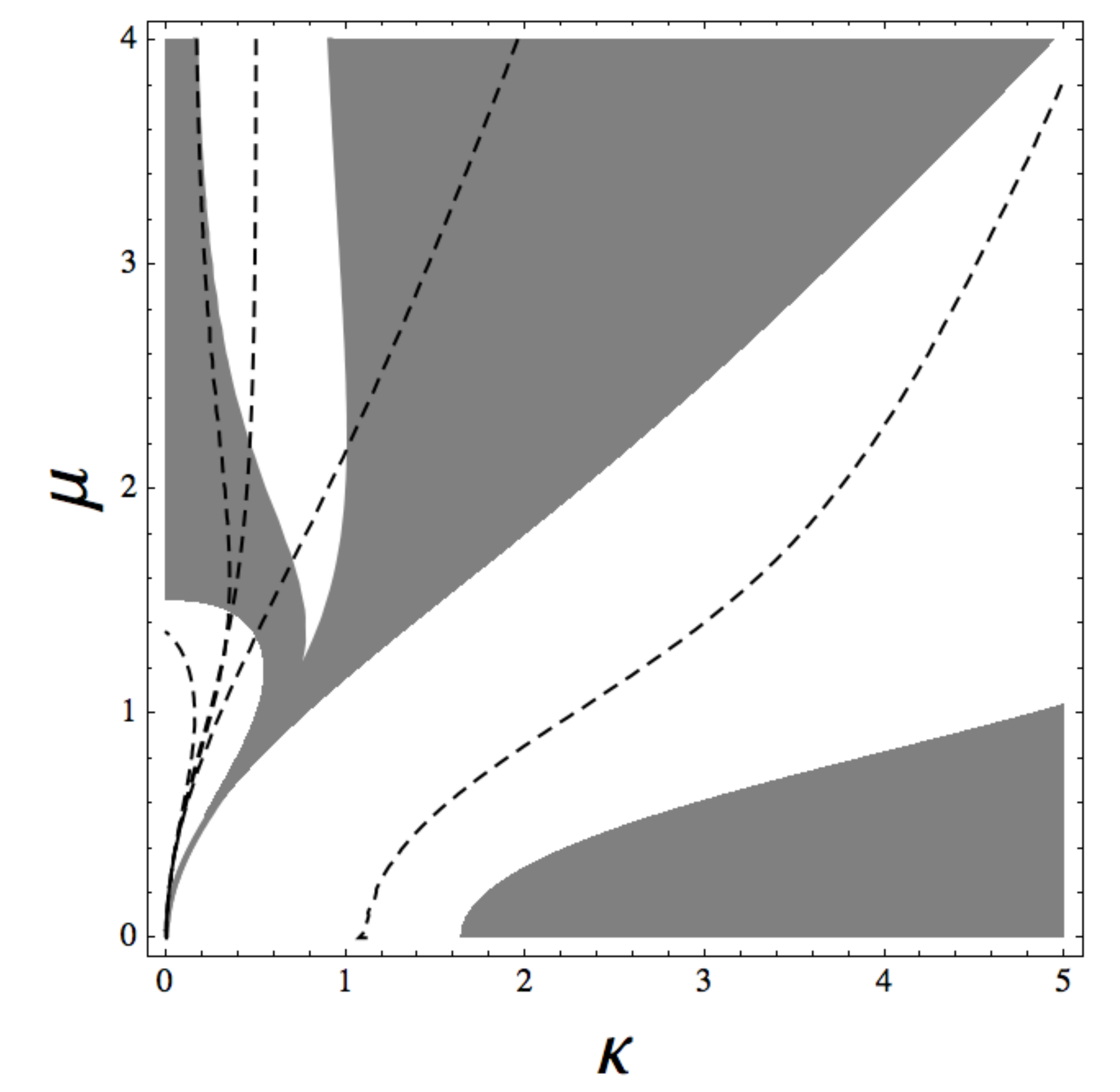} 
\caption{Same as Fig. \ref{instab_omega_m1} for $\Omega=-1$.}
\label{instab_omega_m2}
\end{minipage}\hfill
\begin{minipage}{0.49\textwidth}
\centering
 \includegraphics[width=\linewidth]{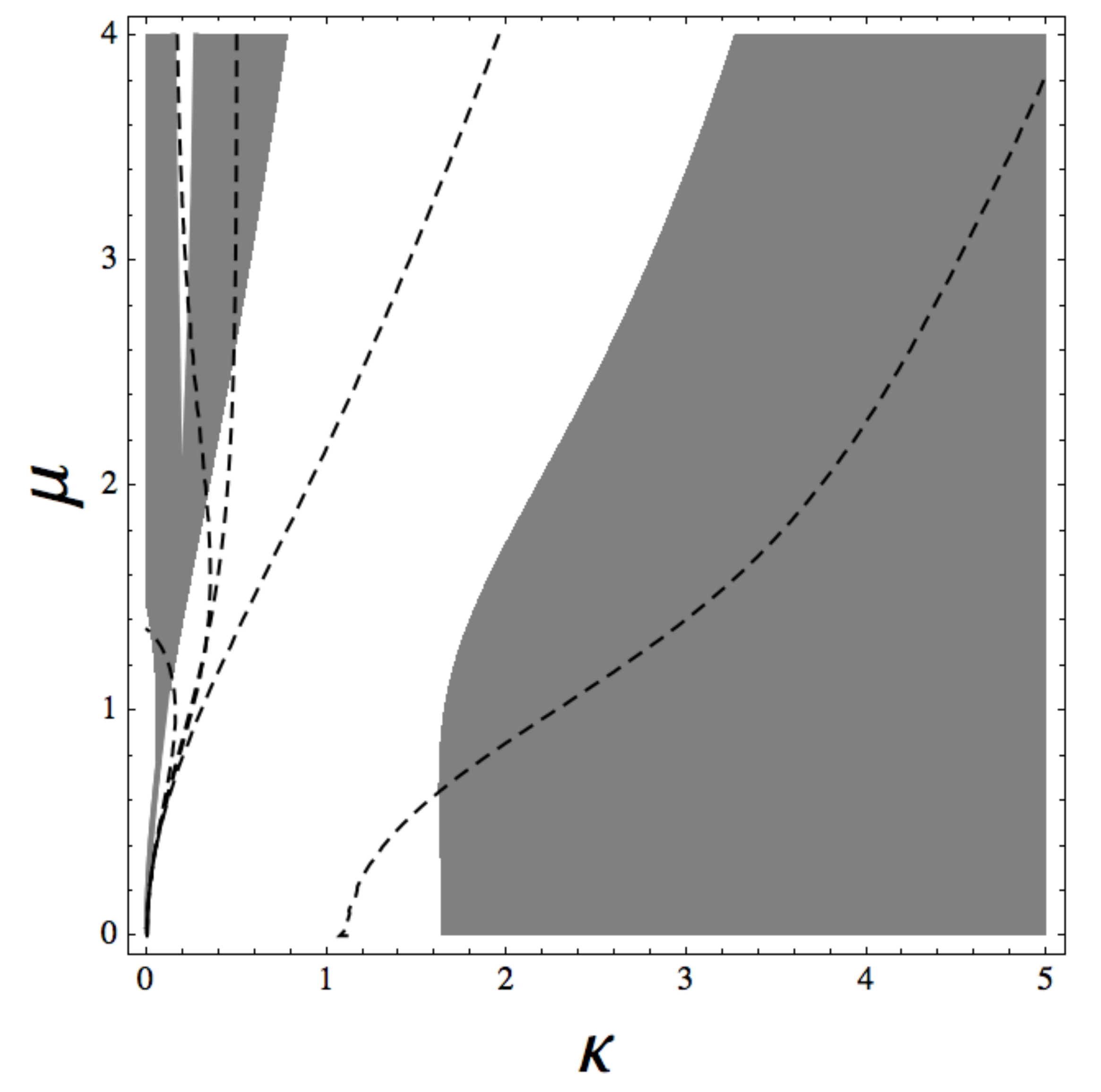}
\caption{Same as Fig. \ref{instab_omega_1} for $\Omega=1$.}
\label{instab_omega_2}
\end{minipage}
\end{figure}
\begin{figure}
\centering
\begin{minipage}{0.49\textwidth}
\centering
\includegraphics[width=\linewidth]{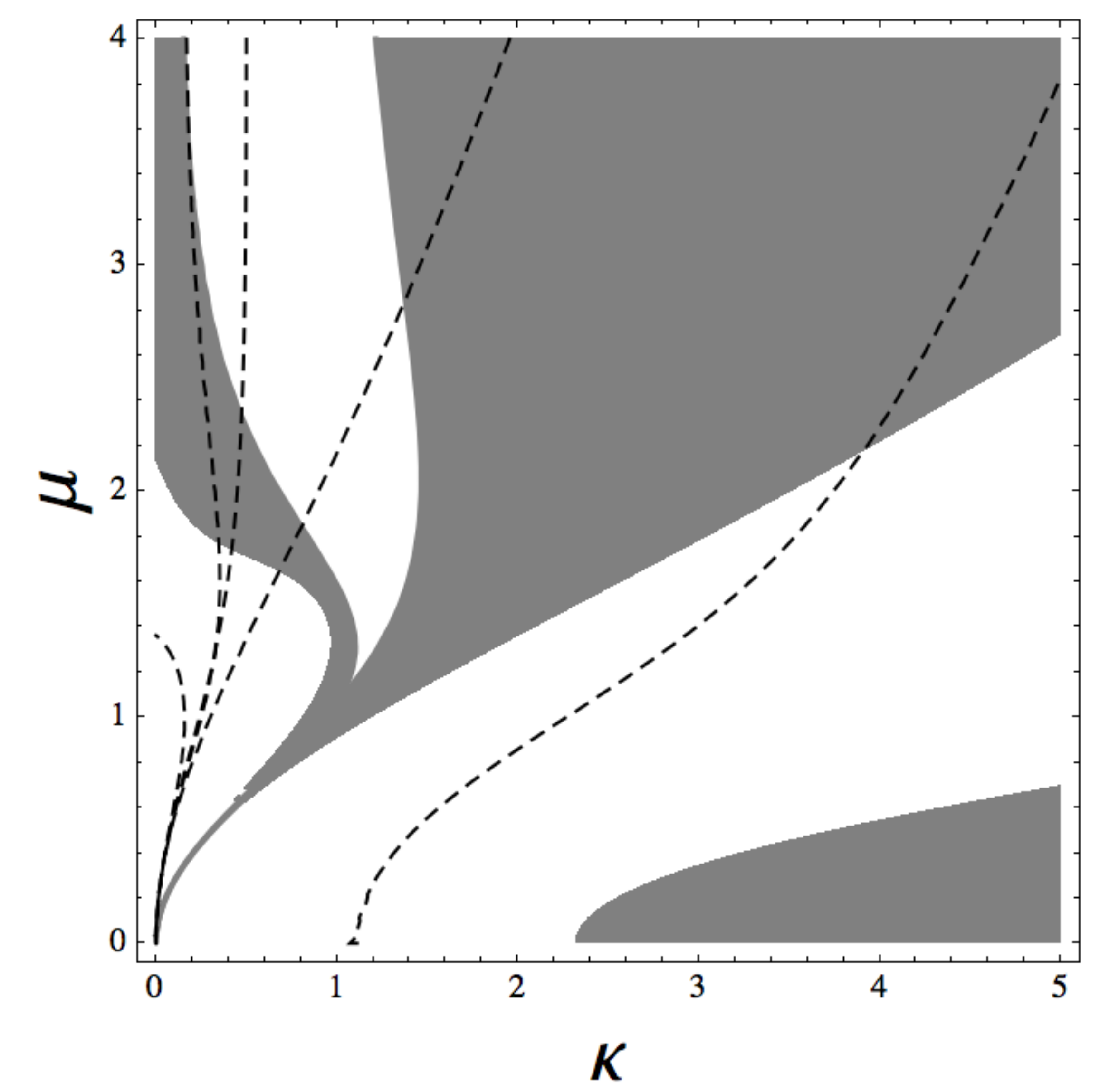} 
\caption{Same as Fig. \ref{instab_omega_m1} for $\Omega=-1.5$.}
\label{instab_omega_m3}
\end{minipage}\hfill
\begin{minipage}{0.49\textwidth}
\centering
 \includegraphics[width=\linewidth]{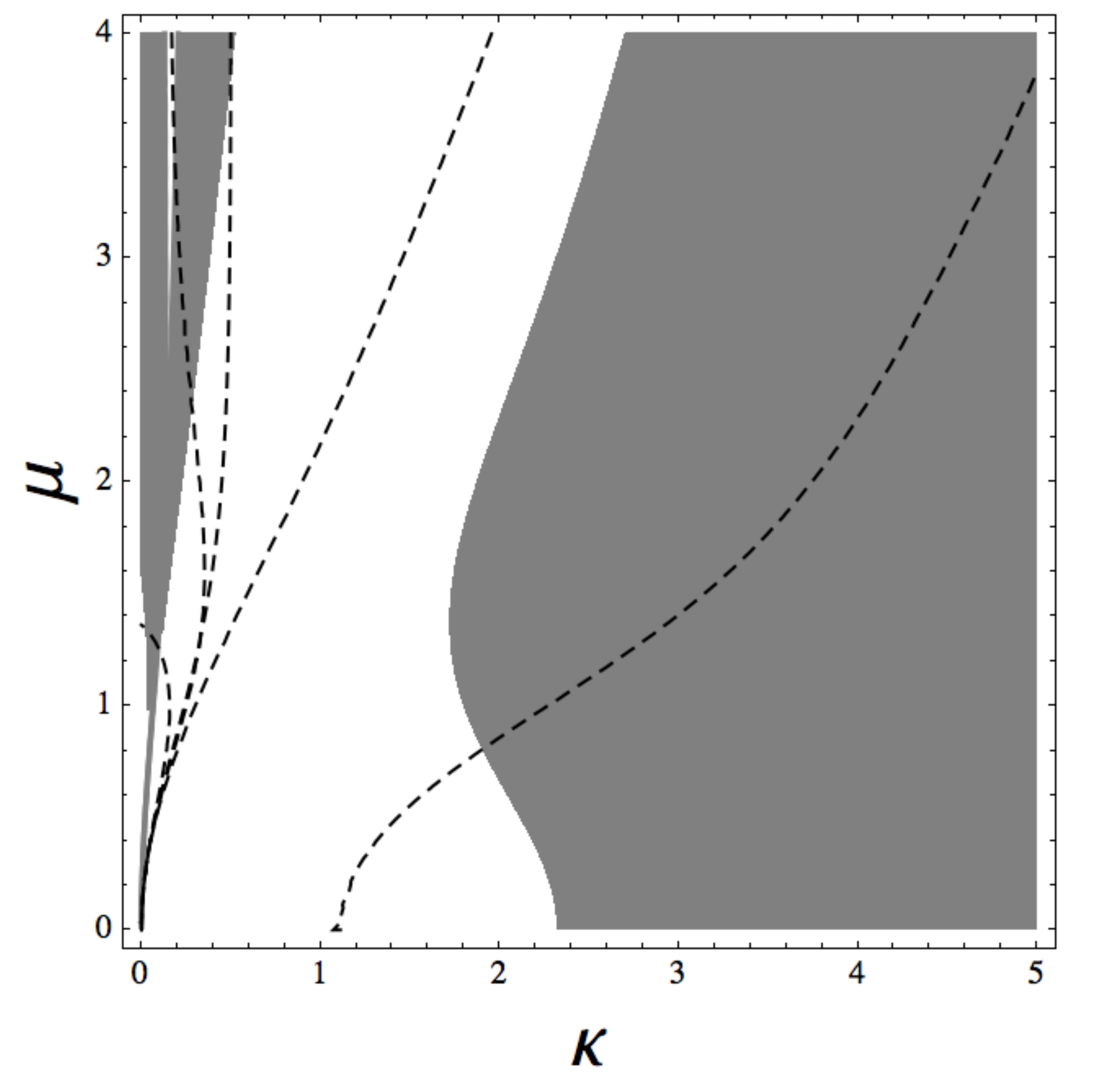}
\caption{Same as Fig. \ref{instab_omega_1} for $\Omega=1.5$.}
\label{instab_omega_3}
\end{minipage}
\end{figure}
\begin{figure}
\centering
\begin{minipage}{0.49\textwidth}
\centering
\includegraphics[width=\linewidth]{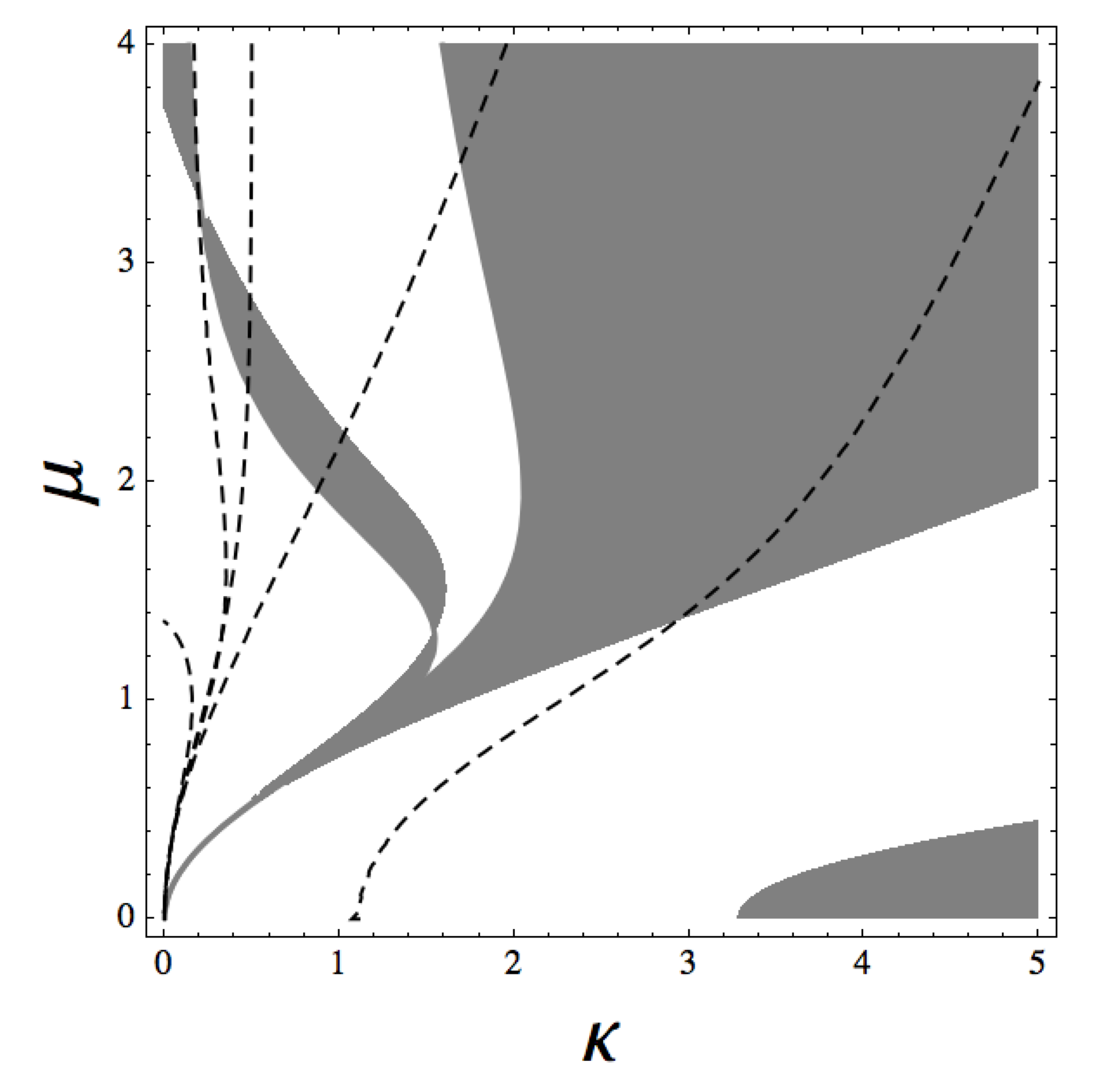} 
\caption{Same as Fig. \ref{instab_omega_m1} for $\Omega=-2$.} 
\label{instab_omega_m4}
\end{minipage}\hfill
\begin{minipage}{0.49\textwidth}
\centering
 \includegraphics[width=\linewidth]{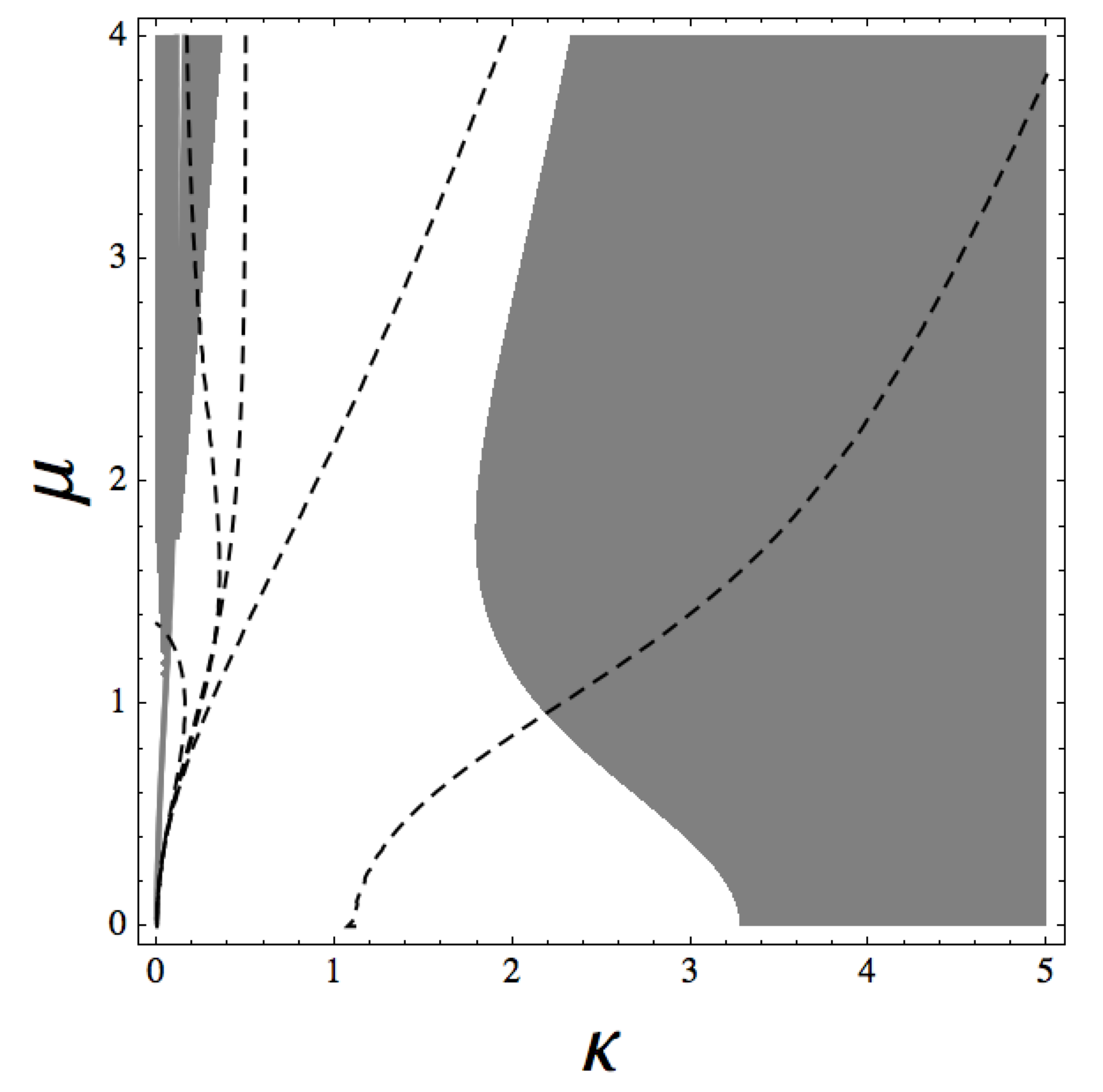}
\caption{Same as Fig. \ref{instab_omega_1} for $\Omega=2$.} 
\label{instab_omega_4}
\end{minipage}
\end{figure}
\section{Conclusion}
A nonlinear Schrödinger equation for capillary-gravity waves in finite depth with a linear shear current has been derived which extends the work of \citet{thomas2012pof}. The combined effect of vorticity and surface tension on modulational instability properties of weakly nonlinear gravity-capillary and capillary wave trains has been investigated. The explicit expressions of the dispersive and nonlinear coefficients are given as a function of the frequency and wavenumber of the carrier wave, the vorticity, the surface tension and the depth. The linear stability to modulational perturbations of the Stokes wave solution of the NLS equation has been carried out. Two kinds of waves have been especially investigated that concerns short gravity waves influenced by surface tension and pure capillary waves. In both cases, vorticity effect is to modify the rate of growth of modulational instability and instability bandwidth. Furthermore, it is shown that vorticity effect modifies significantly the stability diagrams of the gravity-capillary waves. 


\end{document}